\documentclass[12pt,journal,onecolumn,twoside,draftcls]{IEEEtranTCOM}
\usepackage{cite}
\usepackage[centertags]{amsmath}
\usepackage{amssymb}
\usepackage{wasysym}
\usepackage{stmaryrd}
\usepackage{mathrsfs}
\usepackage{flushend}
\DeclareMathAlphabet{\mathpzc}{OT1}{pzc}{m}{it}
\usepackage{algorithmic}
\usepackage{array}
\usepackage{stfloats}
\usepackage{url}
\usepackage{multirow}
\usepackage{algorithm}
\usepackage{color}
\usepackage{bbm}
\usepackage{amsfonts}
\usepackage{amsthm} 
\usepackage{graphicx}
\usepackage{dsfont}
\usepackage{relsize}

\usepackage[tight,footnotesize]{subfigure}

\newcommand{\indic}{\mathlarger{\mathds{1}}}
\newtheorem{defn}{Definition}
\newtheorem{te}{Theorem}
\newtheorem{lema}{Lemma}

\newtheorem{ex}{Example}

\newcommand{\sgn}{\text{sgn}}

\newcommand{\supp}{\mbox{supp}}
\newcommand{\mrm}{\mathrm}

\begin{document}
\title{Finite Alphabet Iterative Decoders, Part I: Decoding Beyond Belief Propagation on BSC}
\author{Shiva~Kumar~Planjery,~\IEEEmembership{Student~Member,~IEEE}, David~Declercq,\\~\IEEEmembership{Senior~Member,~IEEE}, Ludovic~Danjean,			and~Bane~Vasi\'{c},~\IEEEmembership{Fellow,~IEEE}
\thanks{This work was funded by the NSF under Grant CCF-0963726 and Institut Universitaire de France grant. The work was presented in part at the IEEE conferences Information Theory and Applications Workshop 2010, International Symposium on Information Theory 2010, and Information Theory Workshop 2011.} 
\thanks{B. Vasi\'{c} is with the Department
of Electrical and Computer Engineering, University of Arizona, Tucson,
AZ, 85721 USA (e-mail: vasic@ece.arizona.edu).}
\thanks{D. Declercq is with ENSEA/University of Cergy-Pontoise/CNRS UMR 8051, 95014 Cergy-Pontoise, France (email: declercq@ensea.fr).}
\thanks{ S. K. Planjery and L. Danjean are with both the above institutions (email: \{shivap,danjean\}@ece.arizona.edu).}}
%

\maketitle

\begin{abstract}
We introduce a new paradigm for finite precision iterative decoding on low-density parity-check codes over the Binary Symmetric channel. The messages take values from a finite alphabet, and unlike traditional quantized decoders which are quantized versions of the Belief propagation (BP) decoder, the proposed finite alphabet iterative decoders (FAIDs) do not propagate quantized probabilities or log-likelihoods and the variable node update functions do not mimic the BP decoder. Rather, the update functions are maps designed using the knowledge of potentially harmful subgraphs that could be present in a given code, thereby rendering these decoders capable of outperforming the BP in the error floor region. On certain column-weight-three codes of practical interest, we show that there exist 3-bit precision FAIDs  that surpass the BP decoder in the error floor.  Hence, FAIDs  are able to achieve a superior performance at much lower complexity.  We also provide a methodology for the selection of FAIDs that is not code-specific, but gives a set of candidate FAIDs containing potentially good decoders in the error floor region for any column-weight-three code. We validate the code generality of our methodology by providing particularly good three-bit precision FAIDs for a variety of codes with different rates and lengths.

\end{abstract}

\section{Introduction}
\label{sect_intro}
At the heart of modern coding theory lies the fact that low-density parity-check (LDPC) codes \cite{gallager} can be efficiently decoded by message-passing algorithms which are based on the belief propagation (BP) algorithm \cite{pearl}. The BP algorithm operates on a graphical model of a code known as the \emph{Tanner graph}, and computes marginals of functions on the graph. While inference using BP is exact only on loop-free graphs (trees) and exact inference on loopy graphs is hard even under strong restrictions of graphical model topology, the BP still provides surprisingly close approximations to exact marginals on loopy graphs. 

However, the sub-optimality of BP on loopy graphs contributes to the error floor phenomenon of LDPC codes. The error floor is an abrupt degradation in the slope of the error-rate performance in the high signal-to-noise-ratio (SNR) regime, where certain harmful loopy structures, generically termed as {\it trapping sets} \cite{richardsonerrorfloor} present in the Tanner graph of the code, cause the decoder to fail. Although there have been some important works related to devising algorithms that can provide better approximations on the marginals such as the generalized BP algorithms of \cite{yedidia}, these are still too complex for practical use. Moreover, the effects of finite precision that are introduced when decoders are realized in hardware can further contribute to the error floor problem. A glimpse at the iterative decoders developed so far reveals a wide range of decoders of varying complexity. The simple binary message passing algorithms such as the Gallager A/B algorithms \cite{gallager} occupy one end of the spectrum, while the BP lies at the other end. The gamut of decoders filling the intermediate space can simply be understood as the implementation of the BP (and variants) at different levels of precision. 

In this paper, which serves as the first part of our two-part paper series, we introduce a novel approach to the design of finite precision iterative decoders for the Binary Symmetric channel (BSC) which we refer to as {\it finite alphabet iterative decoders} (FAIDs) to signify the fact that the messages belong to a finite alphabet. Some of the key features that clearly distinguish our approach from all the other existing works on finite precision iterative decoders (which shall be discussed in greater detail in the next section) are: 1) the messages are not quantized values of log-likelihoods or probabilities, and 2) the variable node update functions are simple well-defined maps rather than approximations of the update functions used in BP. The maps for variable node update in FAIDs are designed with the goal of increasing the guaranteed error-correction capability by using the knowledge of potentially harmful subgraphs that could be present in any given code, thereby improving the slope of the error floor on the BSC \cite{milos}. Since the variable nodes in the proposed decoders are now equipped to deal with potentially harmful neighborhoods, which is in contrast to BP which treats the loopy Tanner graph as a tree, the proposed decoders are capable of surpassing the floating-point BP in the error floor.  

We restrict our focus to FAIDs for column-weight-three codes. The main reason for this is that such codes, while enabling extremely simple hardware implementations, are notoriously prone to higher error floors especially at moderate to high code rates compared to codes of higher column-weights. Being able to design good yet simple decoders for these codes not only validates our novel approach, but also further ascertains the importance of addressing the error problem from the viewpoint of improving the iterative decoding rather than from the viewpoint of code construction \cite{DungLatinSquare}, as it becomes very difficult to construct such codes to be devoid of certain harmful loopy graphs or with large girth without compromise on the code rate and length.

We also provide a semi-heuristic-based selection method that is not code specific, but gives a set of candidate FAIDs that contains potentially good decoders in the error floor for any given column-weight-three code. The method relies on analyzing the behavior of different FAIDs on a small number of carefully selected subgraphs which are potential trapping sets with errors introduced in the nodes of the subgraph. In order to carry out this analysis, we introduce the notion of {\it noisy trapping set}, which is a generalized notion that takes into account the possible effects of a neighborhood by initializing different possible sets of messages into the trapping set. Using this approach, we obtain a set of particularly good 3-bit precision FAIDs that are each capable of surpassing BP on several column-weight-three codes.


\section{Finite precision iterative decoding and the error floor: prior work}
\label{sec:PriorWork}
There have been several important works related to the design of quantized BP decoders. Early works include the Gallager-E and other finite precision decoders proposed by Richardson and Urbanke \cite{richardson}, and reduced-complexity BP decoders such as the normalized min-sum and offset min-sum decoders proposed by Chen {\it et al.} \cite{chen} and by Fossorier {\it et al.} \cite{Fossorier}. More recent works include the quantized BP decoders proposed by Lee and Thorpe \cite{leethorpe}, and also by Kurkosi and Yagi \cite{kurkoski}. In all of the aforementioned works, the quantization schemes are designed based on optimizing for the best decoding threshold on a given code using the asymptotic technique of density evolution (DE) \cite{richardson} with the primary goal of approaching the performance of the floating-point BP algorithm. Since asymptotic methods are inapplicable to finite length codes as they do not to take their particular structures into account, these decoders designed for the best DE thresholds do not guarantee a good performance on a finite length code especially in the high SNR region. This was also evidenced in \cite{Planjery_ElecLetters_2011,LudoITW2011} where the FAIDs that were chosen for a given code solely based on their DE thresholds were not the best performing FAIDs and in some cases even exhibited high error floors. Moreover, the effects of quantization on the error floor can vary depending on the particular structure of a given code \cite{richardsonerrorfloor,DolecekDec}. 

The error floor problem of LDPC codes has gained significant attention over the past several years with several works addressing the problem from a decoding perspective by proposing modifications to the BP decoding algorithm. Some of the notable works include augmented belief propagation \cite{varnica}, informed dynamic scheduling \cite{Wesel}, multi-stage decoding \cite{ywang}, averaged decoding \cite{laendner}, and use of post-processing to lower the error floors \cite{YHan,zhangerrorfloor}. While all these schemes certainly provided performance enhancements in the error floor, all of them require either a considerable increase in decoding complexity due to the modifications and post-processing or are restricted to a particular code whose structure is well-known. In addition, they do not take finite precision into account and this can drastically affect the performance gains when implemented in hardware due to possible numerical precision issues \cite{DolecekQuantization}.

Therefore, addressing the error floor problem while also taking finite-precision into account in the decoder design is of prime importance especially with emerging applications in communication and data storage systems now requiring very low error-rates and faster processing speeds. In this regard, there are some relevant works worth mentioning. Zhao {\it et al.} in \cite{bannihasemi} proposed several modifications to the offset min-sum decoder while taking quantization into account. Their proposed quantization schemes enabled their offset-min sum decoder to approach performance of floating-point BP algorithm on the Additive White Gaussian Noise channel (AWGNC) with six bits of quantization with possible improvement in the error floor.  Zhang {\it et al.} in \cite{DolecekQuantization} also studied the effects of quantization on the error floors of BP over the AWGNC, and designed schemes that led to substantial error floor reductions when six bits of quantization were used. 

However, using our novel approach, which is different from the above mentioned works (due to the reasons mentioned in Section \ref{sect_intro}), we are able to design simple 3-bit precision decoders that are capable of surpassing the floating-point BP.

\section{Preliminaries}
\label{sec:Preliminaries}
Let $G$ denote the Tanner graph of an $(N,K)$ binary LDPC code $\cal{C}$ of rate $R=K/N$, which consists of the set of variable nodes $V=\{v_1,\cdots,v_N\}$ and the set of check nodes $C=\{c_1,\cdots,c_M\}$. The degree of a node in $G$ is the number of its neighbors in $G$. A code $\cal{C}$ is said to have a regular column-weight $d_v$ if all variable nodes in $V$ of  have the same degree $d_v$.  The set of neighbors of a node $v_i$ is denoted as $\mathcal{N}(v_i)$, and the set of neighbors of node $c_j$ is denoted by $\mathcal{N}(c_j)$.  The girth of $G$ is the length of shortest cycle present in $G$. 

Let $\mathbf{x}=(x_1,x_2,\ldots,x_N)$ denote a codeword of $\cal C$ that is transmitted over the BSC, where $x_i$ denotes the value of the bit associated with variable node $v_i$, and let the vector received from the BSC be $\mathbf{r}=\{r_1,r_2,\ldots,r_N\}$. Let $\mathbf{e}=(e_1,e_2,\ldots,e_N)$ denote the \emph{error pattern} introduced by the BSC such that $\mathbf{r} = \mathbf{x} \oplus \mathbf{e}$, and $\oplus$ is the modulo-two sum. The support of an error vector $\mathbf{e}=(e_1,e_2,\ldots,e_N)$, denoted by $\supp(\mathbf{e})$, is defined as the set of all positions $i$ such that $e_i\neq 0$. Let $\mathbf{y}=(y_1,y_2,\ldots,y_N)$ be the input to the decoder, where each $y_i$ is calculated based on the received value $r_i$. We shall also refer to the values $y_i$ as {\it channel values}. During the analysis of decoders, we shall assume that the all-zero codeword was transmitted. This is a valid assumption since the decoders we consider are symmetric \cite{richardson}. 

A trapping set (TS) denoted by $\mathbf{T}(\mathbf{y})$ (as originally defined in \cite{richardsonerrorfloor}) is a non-empty set of variable nodes that are not eventually corrected for a given decoder input $\mathbf{y}$. If $\mathbf{T}(\mathbf{y})$ is empty, then the decoding is successful. Note that $\mathbf{T}(\mathbf{y})$ will depend on the number of decoding iterations. A common notation used to denote a TS is $(a,b)$, where $a=|\mathbf{T}(\mathbf{y})|$, and $b$ is the number of odd-degree check nodes in the subgraph induced by $\mathbf{T}(\mathbf{y})$. 

Let $\mathcal{T}(a,b)$ denote the topology associated with a $(a,b)$ TS, which is a graph consisting of $a$ variable nodes and $b$ odd-degree check nodes. A TS is said to be elementary if $\mathcal{T}$ contains only degree-one or/and degree-two check nodes. Throughout this paper, we restrict our focus to elementary trapping sets, since they are known to be dominant in the error floor \cite{richardsonerrorfloor,shashiICC}. Also, whenever we refer to a TS, we will implicitly refer to its underlying topological structure $\mathcal{T}$.

\section{Finite alphabet iterative decoders}\label{sec:FAIDs}

We shall now introduce a new type of finite precision decoders which we refer to as FAIDs \cite{Planjery_ElecLetters_2011,PlanjeryITA2010}. An $N_s$-level FAID denoted by $\mathrm{D}$ is defined as a 4-tuple given by $\mathrm{D}=(\mathcal{M},\mathcal{Y},\Phi_v,\Phi_c)$. The finite alphabet $\mathcal{M}$ defined as $\mathcal M = \{-L_s,\ldots,-L_2,-L_1,0,L_1,L_2,\ldots,L_s\}$, where $L_i\in\mathbb{R^{+}}$ and $L_i>L_j$ for any $i>j$, consists of $N_s=2s+1$ levels for which the message values are confined to. The sign of a message $x \in \mathcal M$ can be interpreted as the estimate of the bit associated with the variable node for which $x$ is being passed to or from (positive for zero and negative for one), and the magnitude $|x|$ as a measure of how reliable this value is. 

The set $\mathcal{Y}$ denotes the set of all possible channel values. For FAIDs over the BSC, $\mathcal{Y}$ is defined as $\mathcal{Y}=\{\pm \mrm{C}\}$, where $\mrm{C}\in\mathbb{R^{+}}$, and the value $y_i\in \cal{Y}$ for node $v_i$ is determined by $y_i=(-1)^{r_i}\mathrm{C}$, i.e., we use the mapping $0\rightarrow \mathrm{C}$ and $1\rightarrow -\mathrm{C}$. Note that for the BP and min-sum algorithms (where the messages are log-likelihoods), the decoder input $\mathbf{y}$ is a real-valued vector ($\mathcal{Y}=\mathbb{R}$). Let $m_1,\cdots,m_{l-1}$ denote the extrinsic incoming messages to a node with degree $l$. 

\subsection{Definitions of the update functions $\Phi_v$ and $\Phi_c$}

The function $\Phi_c: \mathcal{M}^{d_c-1} \to \mathcal{M} $ used for update at a check node with degree $d_c$ is defined as
\begin{equation}
\label{eq:minsum}
\Phi_c(m_1,\ldots,m_{d_c-1}) = \left(\prod_{j=1}^{d_c-1}\sgn(m_j)\right) \min_{j \in \{1,\ldots,d_c-1\}}(|m_j|).
\end{equation}
Note that this is the same function used in the min-sum decoder, and hence the novelty in the proposed decoders lies in the definition of the variable node update function $\Phi_v$.  

The function $\Phi_v:\mathcal{Y} \times \mathcal{M}^{d_v-1} \to \mathcal{M} $ is the update function used at a variable node with degree $d_v$, and is defined in closed form as
\begin{equation}
\Phi_v(y_i,m_1,m_2,\cdots,m_{d_v-1})=Q\left(\sum_{j=1}^{d_v-1}m_j+ \omega_i \cdot y_i \right)
\end{equation}
where the function $Q(.)$ is defined below based on a threshold set $\mathscr{T}=\{T_i  :1\leq i\leq s+1 \}$ with $T_i\in\mathbb{R^{+}}$ and $T_i>T_j$ if $i>j$,
and $T_{s+1}=\infty$.
\begin{displaymath}
Q(x)=\Bigg\{ \begin{tabular}{ll}
$\mrm{sgn}(x)L_i$, & $\mrm{if}$ \ $T_{i}\leq |x| < T_{i+1}$ \\
0, & $\mrm{otherwise}$
\end{tabular}
\end{displaymath}

The weight $\omega_i$ is computed using a symmetric function $\Omega:\mathcal{M}^{d_v-1}\to \{0,1\}$. Based on this definition, the function $\Phi_v$ can be classified as a linear-threshold (LT) function or a non-linear-threshold (NLT) function. If $\Omega=1$ (or constant), i.e., if the value of $\omega_i$ is always $1$ (or constant) for all possible inputs of $\Omega$, then $\Phi_v$ is an LT function and a FAID with such a $\Phi_v$ is classified as an LT FAID. Else, $\Phi_v$ is an NLT function and a FAID with such a $\Phi_v$ is an NLT FAID. 

Note that for an LT FAID, $\Phi_v$ takes a linear combination of its arguments and then applies the function $Q$ to determine its output. Therefore, $\Phi_v$ will always output the same value for any possible set of incoming messages for a given $y_i$, if their sum remains the same. For example, for a node with $d_v=3$, $\Phi_v(-\mrm{C},m_1,m_2)=\Phi_v(-\mrm{C},m_3,m_4)$ when $m_1+m_2=m_3+m_4$. This is also a typical property present in existing quantized decoders such as quantized BP and min-sum. 

On the other hand, for an NLT FAID, $\Phi_v$ takes a non-linear combination of its arguments (due to $\Omega$) before applying the function $Q$ on the result. Therefore, $\Phi_v$ can output different values even for distinct sets of incoming messages that have the same sum. For instance, consider a map $\Phi_v$ for a node with $d_v=3$ such that $\Phi_v(-\mrm{C},-L_3,L_3)=0$ and $\Phi_v(-\mrm{C},-L_2,L_2)=-L_1$. In this case, the two distinct sets of incoming messages are $\{-L_3,L_3\}$ and $\{-L_2,L_2\}$, and the sums are zero for both the sets. However, $\Phi_v$ still gives different outputs for each of the sets namely, $0$ and $-L_1$ respectively. Hence these decoders are different from existing quantized message-passing decoders. Note that the function $\Phi_v$ satisfies the following two properties. 

\begin{defn}[Symmetry property]
A FAID is a symmetric decoder if its update function $\Phi_v$ satisfies $\Phi_v(y_i,m_1,...,m_{d_v-1}) = - \Phi_v(-y_i,-m_1,...,-m_{d_v-1})$.
\end{defn}
\begin{defn}[Lexicographic ordering]
A FAID is said to be lexicographically ordered if $\Phi_v$ satisfies $\Phi_v(\mrm{-C},m_1,\ldots,m_{d_v-1}) \geq \Phi_v(\mrm{-C},m'_1,\ldots,m'_{d_v-1})$ $\forall m_i \geq m'_i$, $i\in\{1,\ldots,d_v-1\}$.
\end{defn} 

The property of {\it lexicographic ordering} ensures that the for a given channel value, the output is always non-decreasing with increase in the values of incoming messages. For instance, a map $\Phi_v$ where $\Phi_v(-\mrm{C},L_1,L_2)=L_2$ and $\Phi_v(\mrm{-C},L_2,L_2)=L_1$ is forbidden, since $\Phi_v(-\mrm{C},L_2,L_2)\geq\Phi_v(\mrm{-C},L_1,L_2)$. This is a typical property also present in existing message-passing decoders. 

The decision rule used in FAIDs at the end of each iteration, to determine the bit value corresponding to each node $v_i$, is simply the sign of the sum of all incoming messages plus the channel value $y_i$ (positive implies zero and negative implies one). 

It is evident from the definition that $\Phi_v$ can be uniquely described either by assigning real values to the elements of $\mathcal{M}$, $\mathscr{T}$ and $\mathcal{Y}$, and defining $\Omega$, or by providing a set of constraints which the assigned values can take. As examples, we provide the closed-form description of $\Phi_v$ for a 5-level NLT FAID and a 7-level LT FAID defined for column-weight-three codes. 

\begin{ex}[5-level NLT FAID]\label{5NLTFAID}
The constraints on the values assigned to elements of $\mathcal{M}$ and $\mathscr{T}$ that describe this map are: $\mrm{C}=L_1$, $L_2=3L_1$, $T_1=L_1$, $T_2=L_2$, and the function $\Omega$ is given by $\omega_i =\Omega(m_1,m_2)= 1-\big(\mrm{sign}(m_1) \oplus\mrm{sign}(m_2) \big) \cdot \delta(|m_1|+|m_2|- 2 L_2 )$.  
\end{ex}
\begin{ex}[7-level LT FAID]
\label{7LTFAID}
The constraints on the values assigned to elements of $\mathcal{M}$ and $\mathscr{T}$ that describe this map are: $L_1<\mrm{C}<2L_1$, $L_2=2L_1$, $L_3=2L_2 + \mrm{C}$, and $T_1=L_1$, $T_2=L_2$, and $T_3=L_3-C$, where $\Omega=1$ since it is an LT function. 
\end{ex}
 
Note in Ex. \ref{5NLTFAID}, $\mrm{sign}(x)=1$ if $x<0$, and $\mrm{sign}(x)=	0$ otherwise. Also note that although the rule defined in Ex. \ref{7LTFAID} appears to be similar to a quantized min-sum decoder, the messages in this decoder are not quantized probabilities or log-likelihoods.

\subsection{Describing the maps of $\Phi_v$ as arrays} 
Let us alternatively define $\mathcal{M}$ to be $\mathcal{M}=\{M_1,M_2,\cdots, M_{N_s}\}$ where $M_1=-L_s$, $M_2=-L_{s-1}$,$\cdots$, $M_s=-L_1$, $M_{s+1}=0$, $M_{s+2}=L_2$,$\cdots$, $M_{N_s}=L_s$. Then, $\Phi_v$ can be defined using $d_{v-1}$-dimensional arrays or look-up tables (LUTs) rather than as closed-form functions, which enables simple implementations and also may be more convenient for decoder selection. 

For column-weight-three codes, the map specifying $\Phi_v$ is a simple two-dimensional array defined by $[l_{i,j}]_{1 \leq i \leq N_s,1\leq j \leq N_s}$, where $l_{i,j} \in \mathcal M$, such that $\Phi_v(-\mrm{C},M_i,M_j)=l_{i,j}$ for any $M_i,M_j\in\mathcal{M}$. The values for $\Phi_v(\mrm{C},M_i,M_j)$ can be deduced from the symmetry of $\Phi_v$. Table \ref{LUT244325} shows an example of a $\Phi_v$ defined as an array for a 7-level FAID when $y_i=-\mrm{C}$.

It is easy to see that if the diagonal entries in the array $[l_{i,j}]_{1\leq i \leq N_s, 1\leq j\leq N_s}$ are different,  then $\Phi_v$ must be an NLT function (as discussed previously). However, $\Phi_v$ can still be an NLT function even if all the diagonal entries in the array are the same, as highlighted in the following lemma. 

\begin{figure}[bhtp]
\begin{center}
\includegraphics[angle=0, width=3.5in]{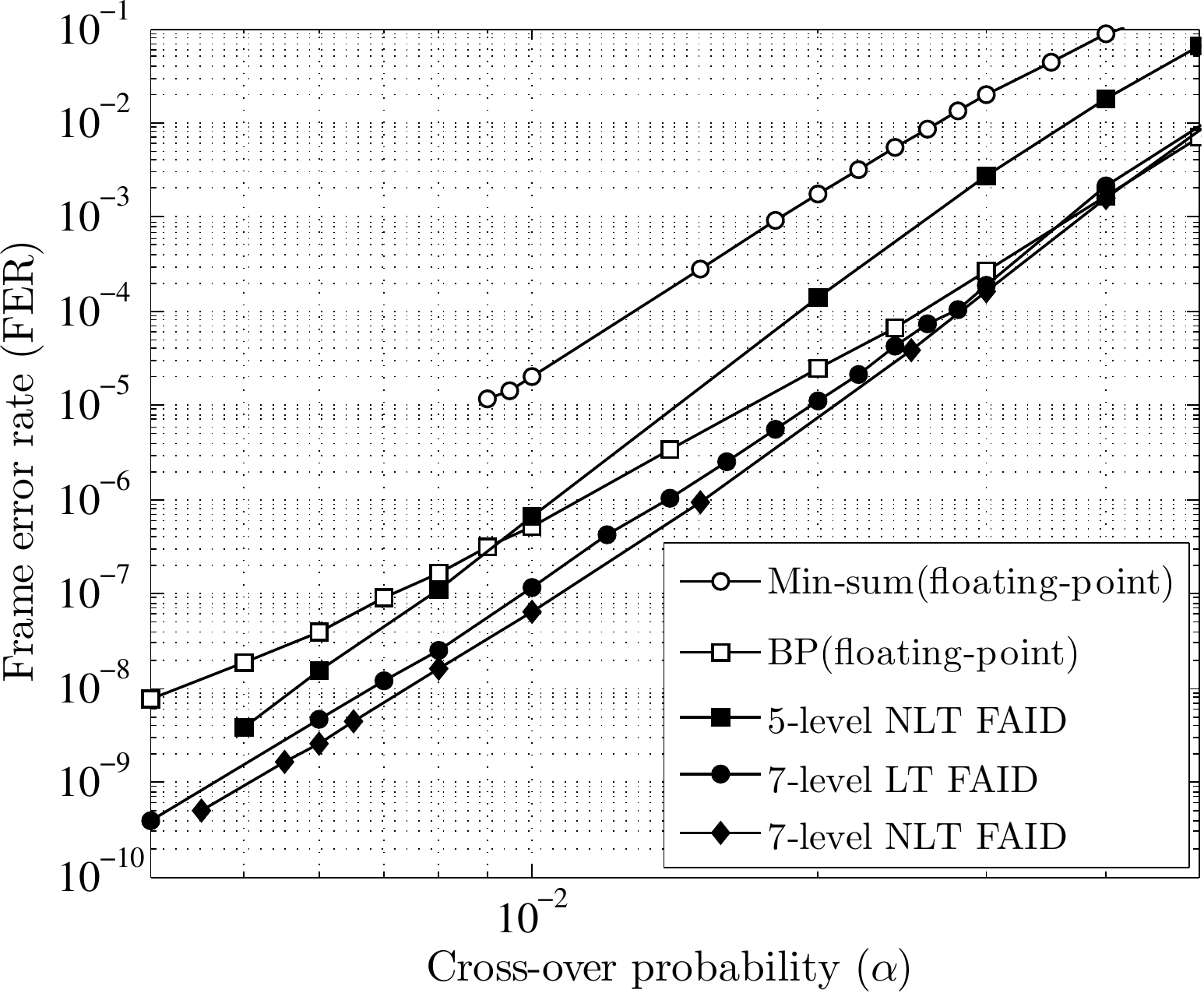}
\caption{Performance comparisons between the floating-point decoders: BP and min-sum , and the 3-bit precision decoders: 5-level NLT, 7-level LT, and 7-level NLT FAIDs on the $(155,64)$ Tanner code.}
\label{TannerPlots}
\end{center}
\end{figure}

\begin{lema}
If $\Phi_v(-\mrm{C},-L_1,L_2)=-L_1$,  $\Phi_v(-\mrm{C},0,L_1)=0$, $\Phi_v(-\mrm{C},0,-L_2)=-L_3$, and $\Phi_v(-\mrm{C},-L_1,-L_1)=-L_2$, then $\Phi_v$ can not be expressed as a linear-threshold function, and hence it is a non-linear-threshold function.
\end{lema}
\IEEEproof See Appendix
\endIEEEproof
Consequently, the map defined by Table \ref{LUT244325} is an NLT function. Fig. \ref{TannerPlots} shows the error-rate performances of the 5-level NLT and 7-level LT FAIDs defined in the two examples, and the 7-level NLT FAID defined by the Table \ref{LUT244325}, along with the floating-point BP and min-sum decoders on the well-known $(155,64)$ Tanner code. All decoders were run for a maximum of 100 iterations. From the plot, we see that all the FAIDs significantly outperform the floating-point BP and min-sum on the code. We will soon provide the methodology used to identify good FAIDs. 

Note that a particular choice of $[l_{i,j}]_{1 \leq i \leq N_s,1\leq j \leq N_s}$ gives rise to a particular $\Phi_v$, and the choice must ensure that the both properties of $\Phi_v$ are satisfied. For the remainder of the paper, we shall restrict our discussion to only class-A FAIDs. A natural question that arises at this point is how many class-A FAIDs exist. This can be easily enumerated by establishing a connection between class-A FAIDs and symmetric plane partitions.

\subsection{Symmetric plane partition representation of $\Phi_v$}

A symmetric plane partition $\pi$ is an array of nonnegative integers $(\pi_{i,j})_{i\geq 1, j\geq 1}$ such that 
$\pi_{i,j} \geq \pi_{i+1,j}$, $ \pi_{i,j} \geq \pi_{i,j+1}$ $\forall i,j\geq 1$, and $\pi_{i,j}=\pi_{j,i}$. If $\pi_{i,j}=0$ $\forall i>r$, $\forall j>s$, and $\pi_{i,j}\leq t$ $\forall i,j$, then the plane partition is said to be \emph{contained} in a box with side lengths $(r,s,t)$.
The value $\pi_{i,j}$ is represented as a box of height $\pi_{i,j}$ positioned at $(i,j)$ coordinate on a horizontal plane. 

\begin{figure}[btp]
\centering
\includegraphics[width=3.5in]{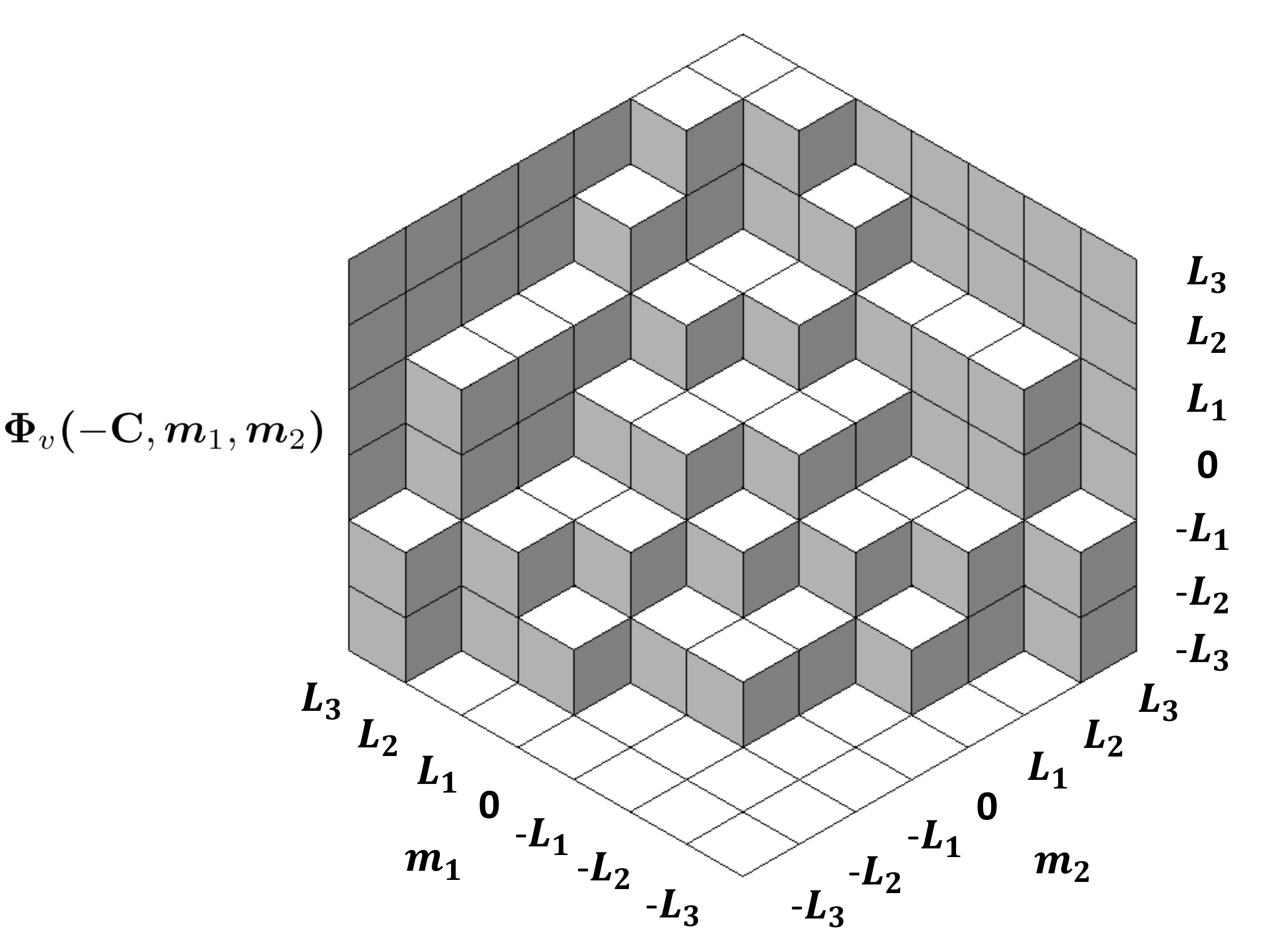}
\caption{A visualization of the plane partition as stacked boxes for the 7-level FAID whose $\Phi_v$ is described in Table \ref{LUT244325}.}
\label{plane7levels}
\end{figure}

Due to the imposition of the lexicographic ordering and symmetry of $\Phi_v$, there exists a bijection between the array $[l_{i,j}]_{1 \leq i\leq N_s,1 \leq j \leq N_s}$ and a symmetric plane partition contained in a $(N_s \times N_s \times N_s-1)$ box, where each $\pi_{i,j}$ is determined based on $l_{i,j}$. Fig. \ref{plane7levels} shows the visualization of a plane partition corresponding to $\Phi_v$ of the 7-level FAID defined in Table \ref{LUT244325}.
Kuperberg in \cite{Kuperberg_symmetries_of_plane_partitions} gave an elegant formula for the enumeration of symmetric plane partitions contained in a box, and we can directly utilize this for the enumeration of class-A $N_s$-level FAIDs as well.

\begin{te}[Number of Class-A $N_s$-level FAID]\label{number_of_FAID}
The total number $K_A(N_s)$ of symmetric lexicographically ordered $N_s$-level FAIDs is given by
\begin{displaymath}
K_A(N_s)=\frac{H_2(3N_s)H_1(N_s)H_2(N_s-1)}{H_2(2N_s+1)H_1(2N_s-1)}
\label{eq:no_decoders}
\end{displaymath}
where $H_k(n) = (n-k)!\,(n-2k)!\,(n-3k)! \ldots\;$ is the staggered hyperfactorial function.
\end{te}
\IEEEproof The proof of the theorem follows from the bijection between the map $\Phi_v$ of a class-A FAID and a symmetric boxed plane partition.
\endIEEEproof

The total number of class-A FAIDs for $N_s=5$ and $N_s=7$ levels are 28,314 and 530,803,988 respectively. Henceforth, we shall restrict our discussion to only class-A FAIDs.

\section{Selection of finite alphabet iterative decoders}\label{sec:SelectionOFFAIDs} 
It is evident from the previous section that identifying particularly good FAIDs from the set of all possible class-A FAIDs is highly non-trivial since the number of such FAIDs is still large. We now describe a general approach that can be used to identify a subset of candidate $N_s$-level FAIDs, one or several of which are potentially good for any column-weight-three code. Our main aim behind this approach is to restrict the choice of FAIDs to a possibly small subset containing good candidates. Given a particular code, it would then be feasible to identify the best performing FAID from this subset by using brute-force simulation or emulation or some other technique on the code. Moreover, since the performance of a FAID on a given code depends on its structure, the goal of identifying several candidate FAIDs is more realistic than identifying a single good FAID, and allows for devising a selection method that is not code-specific. Another important objective of our approach is to ensure that any FAID belonging to this subset is capable of surpassing BP in the error floor not just on a single code but on several codes. 

The approach we use relies on the knowledge of potentially harmful subgraphs that could be trapping sets for traditional iterative decoders when present in a given code. The candidate FAIDs are chosen by analyzing their behavior on each of these subgraphs with errors introduced in them. We will first introduce some important notions that form the basis of our approach, and then subsequently present a methodology for FAID selection for column-weight-three codes. 

\subsection{Critical number and isolation assumption}
The notion of {\it critical number} associated with a TS of type $\mathcal{T}(a,b)$ was originally introduced for Gallager-A/B algorithms on the BSC \cite{shashiITpaper}. It is computed by analyzing the Gallager-A/B decoding on errors contained in the topology $\mathcal{T}$ that is present in a code, assuming that all nodes outside the topology are initially correct. It provides a measure of how harmful a TS is, and hence, this notion is not only useful for predicting the error floor performance \cite{shashiICC} but also for determining the harmful subgraphs that should be avoided in the code designs.

In order to be able to extend the notion of critical number for FAIDs, we introduce the notion of {\it isolation assumption} \cite{planjeryISIT2010} which is used to analyze the decoder on a potential TS $\mathcal{T}(a,b)$. Under this assumption, the neighborhood of the TS is such that the messages flowing into the TS from its neighborhood are not in any way influenced by the messages flowing out of the TS. Therefore, the messages flowing into the TS can be computed while completely disregarding the neighborhood \cite[Theorem 1]{planjeryISIT2010}. We shall now precisely define this notion. 

Let $\mathscr{T}_{i}^k(G)$ denote the computation tree corresponding to an iterative decoder on $G$ enumerated for $k$ iterations with node $v_i\in V$ as its root. A node $w\in \mathscr{T}_{i}^k(G)$ is a {\it descendant} of a node $u\in \mathscr{T}_{i}^k(G)$ if there exists a path starting from node $w$ to root $v_i$ that traverses through node $u$. 

\begin{defn}
[Isolation assumption] Let $H$ be a subgraph of $G$ induced by $P\subseteq V$ with check node set $W\subseteq C$. The computation tree $\mathscr{T}_i^k(G)$ with the root $v_i\in P$ is said to be isolated if for any node $u\notin P\cup W$ in $\mathscr{T}_i^k(G)$, $u$ does not have any descendant belonging to $P\cup W$. If $\mathscr{T}_i^k(G)$ is isolated $\forall v_i \in P$, then $H$ is said to satisfy the isolation assumption in $G$ for $k$ iterations.
\end{defn}
\noindent{\it Remark:} The above definition is a revised version of the one given in \cite{planjeryISIT2010}. 

Note that the isolation assumption is weaker than Gallager's independence assumption as explained in \cite{planjeryISIT2010}. The {\it critical number} can now be defined in the framework of FAIDs.
\begin{defn}
The {\it critical number} of a FAID denoted by $\mrm{D}$ on a subgraph $H$ is the smallest number of errors introduced in $H$ for which $\mrm{D}$ fails on $H$ under the isolation assumption.
\end{defn}
\noindent{\it Remark}: We set the critical number to $\infty$ if $\mrm{D}$ corrects all possible error patterns on $H$.

The critical number can now be used as possible parameter for decoder selection where a decoder is chosen to maximize the critical number on a given TS(s). In principle, one could consider a database of potential TSs that are generated either through analytical or empirical evaluations of traditional decoders such as BP and min-sum on several different codes, and then select a FAID based on its critical numbers on all these TSs. 

However, the isolation assumption of a TS typically does not hold in an actual code for more than few iterations and hence the critical number may
not reflect the true error-correction capability of the FAID on a code containing the TS. This is especially true for TSs of small sizes. Therefore, unless a very large database of TSs is considered or unless TSs with large sizes are considered such that isolation assumption holds for many more iterations (as done in \cite{Declercq_ISTC_2010} where the considered TSs are the weight-20 codewords), the strategy will remain ineffective. This motivates the need for a new notion that considers to an extent the influence of the neighborhood.

\subsection{Noisy trapping sets and noisy critical numbers}\label{sec:Noisy}
Let us consider a harmful topology $\mathcal{T}(a,b)$ that has been identified as a potential TS on a given code. We introduce the notion of {\it initialization vector} which allows us to partially capture the influence of its arbitrary (unknown) neighborhood during the analysis of a FAID on the $\mathcal{T}$.

\begin{defn}
An initialization vector on a TS $\mathcal{T}(a,b)$ is defined as a vector $\mathbf{\Theta}=(\theta_{1},...,\theta_{b})$ where $\theta_{i} \in \mathcal M$, such that during the message passing of a FAID on $\mathcal{T}$, the message passed by the ${i}^{th}$ degree-one check node in any iteration is $\theta_{i}$. The TS $\mathcal{T}(a,b)$ is said to be {\it initialized} by such a vector and is referred to as a noisy trapping set. 
\end{defn}

A FAID can now be analyzed by introducing errors into the variable nodes of the TS $\mathcal{T}(a,b)$ and passing messages iteratively on the edges of $\mathcal{T}$  under a given initialization vector. Note that the initialization vector on a TS is carried out only through the degree-one check nodes, and also that the initialization vector $\mathbf{\Theta}$ is not iteration-dependent.

\begin{figure}[bhtp]
\centering
\includegraphics[width=3in]{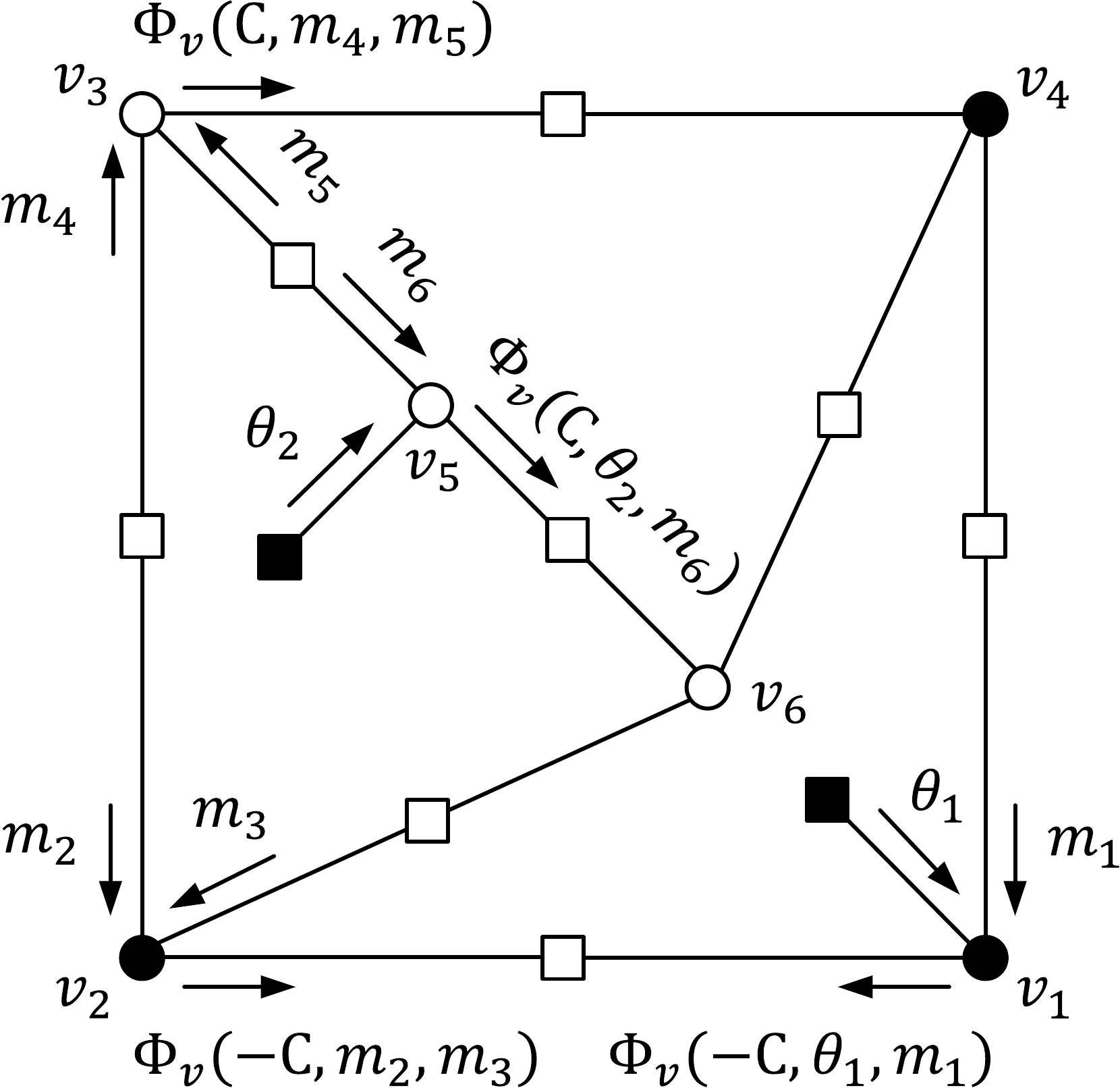}
\caption{An example of a noisy $\mathcal{T}(6,2)$ initialized by a vector $\mathbf{\Theta}$.}
\label{NoisyTSExample}
\end{figure}

As an example, Fig. \ref{NoisyTSExample} depicts how a FAID is analyzed for a three-error pattern on a $\mathcal{T}(6,2)$ initialized by a vector $\mathbf{\Theta}=(\theta_1,\theta_2)$. A $\CIRCLE$ denotes a variable node initially wrong ($v_1$, $v_2$, and $v_4$) and a $\Circle$ denotes a node initially correct ($v_3$, $v_5$, and $v_6$). A $\Box$ denotes a degree-two check node and a $\blacksquare$ denotes a degree-one check node. Initially all the messages passed by all nodes except the degree-one check nodes are set to zero. Then the messages are iteratively updated using the maps $\Phi_v$ and $\Phi_c$ by treating the topology $\mathcal{T}$ as if it were the Tanner graph of a code but with the exception that a degree-one check node sends $\theta_1$ (or $\theta_2$) to its neighbors in all iterations of the message passing. The message update on a single edge from a variable node is shown in the figure for each of the nodes $v_1$, $v_2$, $v_3$ , and $v_5$ ($v_4$ and $v_6$ are similar to $v_2$ and $v_3$ respectively). Note that the messages $m_1,m_2,\ldots,m_6$ denote the extrinsic incoming messages to these nodes. 

Let $N_I$ denote the maximum number of iterations allowed for message passing under a particular FAID on TS $\mathcal{T}(a,b)$. We examine whether an error pattern is corrected by the FAID within $N_I$ iterations under a given initialization vector on the TS $\mathcal{T}(a,b)$. 

Our main intuition for defining such a notion is as follows. Let us consider a code whose graph $G$ contains a subgraph $H$ that is isomorphic to the topology $\mathcal{T}(a,b)$. Assume that a particular FAID is being used for decoding an error pattern where some (or all) of the variable nodes in $H$ are initially in error and the nodes outside $H$ are initially correct. During each iteration of decoding, different possible messages belonging to $\mathcal{M}$ will be passed into the nodes of $H$ from outside of $H$ depending on its neighborhood. The initialization vector can be considered as a possible snapshot of the messages entering $H$ through its check nodes in some arbitrary iteration, and different initializations represent the different possible influences that the neighborhood of $H$ can have. Therefore, analyzing the FAID under different initializations on a given $\mathcal{T}$ can provide a good indication of its error correction capability on a code whose graph contains $H$.

Although the initialization vector should ideally be iteration-dependent and include all messages passed to all check nodes of $\mathcal{T}(a,b)$ from outside of $\mathcal{T}(a,b)$, this would make analyzing a FAID on $\mathcal{T}(a,b)$ computationally intractable. Therefore we only include constant values that are passed to degree-one check nodes into the initialization vector. We now define the notion of {\it noisy critical number} which is an extension of the notion of critical number for FAIDs. 

\begin{defn}
The noisy critical number of a FAID $\mrm{D}$ under an initialization vector $\mathbf{\Theta}$ on a TS $\mathcal{T}(a,b)$ is the smallest number of errors introduced in $\mathcal{T}(a,b)$ for which $\mrm{D}$ fails on $\mathcal{T}(a,b)$. 
\end{defn}

By determining the noisy critical number under every possible initialization vector $\mathbf{\Theta}\in \mathcal{M}^b$ on the TS $\mathcal{T}(a,b)$, a vector of noisy critical numbers, referred to as {\it noisy critical number vector} (NCNV), can be obtained for a particular FAID. Let $N_{\Theta}$ denote the number of all possible initialization vectors, i.e., $N_{\Theta}=|\mathcal{M}^{b}|$. The NCNV of a FAID denoted by $\mrm{D}$ on a given TS $\mathcal{T}(a,b)$ is given by $\mathscr{N}_{\mrm{D}}(\mathcal{T}(a,b),N_I)=(\zeta_1,\zeta_2,\ldots,\zeta_{N_\Theta})$, where $\zeta_i$ is the noisy critical number determined under a initialization vector $\mathbf{\Theta_i}\in \mathcal{M}^b$ on TS $\mathcal{T}(a,b)$ with $N_I$ being the maximum number of decoding iterations. The NCNV can now be used as a parameter for decoder selection. 

\subsection{Choice of trapping sets for decoder selection}
Since our approach for identifying good FAIDs relies on determining the NCNVs of FAIDs on different trapping sets, the first step in the decoder selection is to carefully select the harmful topologies that should be considered for the analysis. The selected trapping sets should be topologies that are known to exist in practical high-rate codes with dense graphs and are regarded as relatively harmful for existing iterative decoders. Also the trapping sets used should have notable differences in their topological structures, so that the candidate FAIDs identified from the analysis are more likely to be good on several codes rather than just on a single code. 

We use the trapping set ontology (TSO) \cite{ontology} to determine which harmful topologies to consider. The TSO is a systematic hierarchy of trapping sets that is based on their topological relations, and it is specified in the form of a {\it parent-child} relationship between the trapping sets. A trapping set $\mathcal{T}_1$ is said to be a parent of a trapping set $\mathcal{T}_2$ if $\mathcal{T}_2$ contains $\mathcal{T}_1$. For the decoder selection, the trapping sets are chosen such that they do not have many common parents, and that most of the parents (graphs of smaller size) in the TSO are considered. For simplicity, we ensure that all the trapping sets selected have the same value of $b$, so that the NCNVs determined from different trapping sets all have the same dimension.

\subsection{Decoder domination}
Having selected the harmful topologies, the next step in the decoder selection is to determine and be able to compare the NCNVs of different FAIDs on all the selected TSs. We introduce the notion of {\it decoder domination} in order to compare the NCNVs of different FAIDs.

Let the set of chosen TSs for the analysis of FAIDs be denoted by $\Lambda=\{\mathcal{T}_1,\mathcal{T}_2,\ldots,\mathcal{T}_{N_\Lambda}\}$ with cardinality $N_\Lambda$. Let $\mathcal{F}=\{\mrm{D}_1,\ldots,\mrm{D}_{N_\mathcal{F}}\}$ denote the set of class-A $N_s$-level FAIDs considered for possible decoder selection with cardinality $N_{\mathcal{F}}$. Let $\mathscr{N}_{\mrm{D}_k}(\mathcal{T}_j,N_I)$ denote the NCNV of a FAID $\mrm{D}_k\in \mathcal{F}$ determined on a TS $\mathcal{T}_j\in \Lambda$, and let $\mathscr{N}_{\mrm{D}_k}^{(i)}(\mathcal{T}_j,N_I)$ denote the $i^{th}$ component of the NCNV, i.e., $\mathscr{N}_{\mrm{D}_k}^{(i)}(\mathcal{T}_j,N_I)=\zeta_i$.

A FAID $\mrm{D}_k$ is said to dominate a FAID $\mrm{D}_l$ for a given initialization vector $\mathbf{\Theta_i}$,  if 
\begin{equation}
\mathscr{N}_{\mrm{D}_k}^{(i)}(\mathcal{T}_j,N_I)\geq \mathscr{N}_{\mrm{D}_l}^{(i)}(\mathcal{T}_j,N_I) \ \ \forall j\in\{1,2,\ldots,N_{\Lambda}\}
\end{equation}
In other words, $\mrm{D}_k$ dominates $\mrm{D}_l$ under a given initialization vector $\mathbf{\Theta_i}$ if the noisy critical number of $\mrm{D}_k$ is not less than the noisy critical number of $\mrm{D}_l$ on all the TSs in $\Lambda$. 

The number of initialization vectors under which $\mrm{D}_k$ dominates $\mrm{D}_l$ is denoted by $\tilde{n}(\mrm{D}_k,\mrm{D}_l)$ and is given by
\begin{equation}
\tilde{n}(\mrm{D}_k,\mrm{D}_l)=\sum_{i=1}^{N_\Theta}\prod_{j=1}^{N_\Lambda}{\indic}\left(\mathscr{N}_{\mrm{D}_k}^{(i)}(\mathcal{T}_j,N_I)\geq \mathscr{N}_{\mrm{D}_l}^{(i)}(\mathcal{T}_j,N_I)\right)
\end{equation}
where $\indic$ is the indicator function that outputs a one when the condition in its argument is true and zero otherwise. 

If $\tilde{n}(\mrm{D}_k,\mrm{D}_l)\geq \tilde{n}(\mrm{D}_l,\mrm{D}_k)$, then $\mrm{D}_k$ is said to dominate $\mrm{D}_l$ with {\it domination strength} $\tilde{n}(\mrm{D}_k,\mrm{D}_l)-\tilde{n}(\mrm{D}_l,\mrm{D}_k)$. For simplicity we shall use the symbol $\rhd$ to denote domination, i.e., $(\mrm{D}_k \rhd \mrm{D}_i)=1$ implies that $\mrm{D}_k$ dominates $\mrm{D}_i$.

\subsection{Methodology for selection: a general approach}

For a given value of $N_s$, a methodology for identifying good $N_s$-level FAIDs can now be devised based on the notions of decoder domination and the NCNVs. We remind the reader that the main goal of our approach is to be able to identify a small subset of candidate $N_s$-level FAIDs, where each candidate FAID is potentially good on several codes. Let this small subset of selected FAIDs be denoted by $\mathcal{F}_c$. Ideally, if a candidate FAID could be selected solely based on how it dominates all the other FAIDs in $\mathcal{F}$, then one could possibly obtain an ordering of the FAIDs in $\mathcal{F}$ in terms of their dominance and conclude as to which ones are more likely to be good on a given code containing one or more of the TSs in $\Lambda$. Unfortunately, we have found that such an ordering does not exist since there can be many FAIDs that dominate a particularly good FAID (known a priori to be good) and yet perform poorly on certain codes.

Therefore, without going into the details, we shall describe a general approach for selection that utilizes pre-determined small sets of good FAIDs and bad FAIDs denoted by $\mathcal{F}_g$ and $\mathcal{F}_b$ respectively. The set $\mathcal{F}_g$ consists of $N_s$-level FAIDs that are known a priori to have good error floor performance on several codes of different rates and possibly containing different TSs. The set $\mathcal{F}_b$ consists of $N_s$-level FAIDs that were found to perform well on one particular code but perform poorly on other codes. We regard FAIDs in $\mathcal{F}_b$ to be bad since our goal is to identify FAIDs that are capable of surpassing BP on several codes. 

We then evaluate whether a particular FAID $\mrm{D}_k\in \mathcal{F}$ dominates or is dominated by the FAIDs in the sets $\mathcal{F}_g$ and $\mathcal{F}_b$. By using the sets $\mathcal{F}_g$ and $\mathcal{F}_b$ to compare with, we are inherently trying to select FAIDs whose NCNVs have characteristics similar to the NCNVs of FAIDs in $\mathcal{F}_g$ but dissimilar to the NCNVs of the FAIDs in $\mathcal{F}_b$. Therefore, we define a cost function, $\mathscr{C}_{\tilde{n}}$, that is based on domination strengths, and whose value determines whether the FAID $\mrm{D}_k$ should be accepted for inclusion into $\mathcal{F}_c$. We have observed that it is crucial for a candidate FAID to dominate most (or all) FAIDs in $\mathcal{F}_g$ and also not be dominated by most (or all) FAIDs in $\mathcal{F}_b$ for it to be considered potentially good. 
This is reflected in the cost function $\mathscr{C}_{\tilde{n}}$ defined below.
\begin{equation}
\label{eq:domination}
\begin{split}
\mathscr{C}_{\tilde{n}}(\mrm{D}_k)= & 
\sum_{\forall \ \mrm{D}_i\in\mathcal{F}_g, \ (\mrm{D}_k\rhd\mrm{D}_i)=1}\big(\tilde{n}(\mrm{D}_k,\mrm{D}_i)-\tilde{n}(\mrm{D}_i,\mrm{D}_k)\big)+ \sum_{\forall \ \mrm{D}_j\in\mathcal{F}_b, \ (\mrm{D}_k\rhd\mrm{D}_j)=1}\big(\tilde{n}(\mrm{D}_k,\mrm{D}_j)-\tilde{n}(\mrm{D}_j,\mrm{D}_k)\big)\\ & - 
\sum_{\forall \ \mrm{D}_i\in\mathcal{F}_g, \ (\mrm{D}_i\rhd\mrm{D}_k)=1}\big(\tilde{n}(\mrm{D}_i,\mrm{D}_k)-\tilde{n}(\mrm{D}_k,\mrm{D}_i)\big) -
\sum_{\forall \ \mrm{D}_j\in\mathcal{F}_b, \ (\mrm{D}_j\rhd\mrm{D}_k)=1}\big(\tilde{n}(\mrm{D}_j,\mrm{D}_k)-\tilde{n}(\mrm{D}_k,\mrm{D}_j)\big)
\end{split}
\end{equation}

The value of the cost function $\mathscr{C}_{\tilde{n}}$ is compared to a threshold $\tau$. If $\mathscr{C}_{\tilde{n}}(\mrm{D}_k)\geq \tau$, then the FAID $\mrm{D}_k$ is selected as a candidate to be included in $\mathcal{F}_c$, else it is rejected. The cardinality of $\mathcal{F}_c$ depends on $\tau$ since a smaller $\tau$ accepts more FAIDs and a larger $\tau$ accepts less FAIDs. The choice of $N_I$ also plays a role and should generally be chosen to be small (5 to 10 iterations). 

Note that the approach we have presented in this paper is slightly different from the one proposed in \cite{LudoITW2011}. In \cite{LudoITW2011}, the selection algorithm assumes it has no a priori knowledge on the sets $\mathcal{F}_g$ and $\mathcal{F}_b$, and then tries to progressively build the sets before using them to identify good candidate FAIDs. By instead utilizing pre-determined sets of $\mathcal{F}_g$ and $\mathcal{F}_b$ in our approach in this paper, we have found that the selection procedure is greatly improved and we were able to obtain much better sets of candidate FAIDs $\mathcal{F}_c$ (in terms of their error floor performance). Note however that the approach of \cite{LudoITW2011} is still applicable to the selection method presented in this paper as it could still be used as an initial step for determining the sets $\mathcal{F}_g$ and $\mathcal{F}_b$.
 
Using our methodology, we were able to derive a set of good candidate 7-level FAIDs (which are 3-bit precision decoders) for column-weight-three codes. On a variety of codes of different rates and lengths, particularly good 7-level FAIDs chosen from $\mathcal{F}_c$ all outperformed the BP (floating-point) in the error floor. Moreover, the loss in the waterfall compared to BP was found to be very reasonable. The numerical results to support this statement are provided in the next section. Another interesting remark related to our selection procedure that we have found is that, although the DE threshold values were not at all used as parameters in the selection of FAIDs, the candidate FAIDs that we obtained in set $\mathcal{F}_c$ were all found to have fairly good DE thresholds. 

\section{Numerical results}\label{sec:Results}
Earlier in Section \ref{sec:FAIDs}, we demonstrated the capability of  5-level and 7-level FAIDs to outperform BP in the error floor on the $(155,64)$ Tanner code. We now provide additional numerical results on the BSC to further illustrate the efficacy of FAIDs on column-weight-three codes of higher practical interest and validate our approach for decoder selection. The three codes used for the simulations were chosen to cover a broad variety of LDPC codes in terms of rate, length, and structure. They are: 1) an $R=0.751$ $(2388,1793)$ structured code based on latin squares, 2) an $R=0.5$ $(504,252)$ code, and
3) an $R=0.833$ $(5184,4322)$ quasi-cyclic code.

The $(2388,1793)$ structured code with girth-8 was designed using the method of Nguyen {\it et. al} \cite{DungLatinSquare}, which is based on latin squares and avoids certain harmful trapping sets in the code design. The $R=0.5$ $(504,252)$ code with girth-8 was designed using the progressive edge-growth (PEG) method of \cite{PEGpaper} while ensuring that it contains no $(5,3)$ TS (see \cite{ontology} for the topology). The $(5184,4322)$ quasi-cyclic code is a high-rate girth-8 code with a minimum distance of 12. 

\begin{figure}[bhtp]
\begin{center}
\includegraphics[angle=0, width=3.5in]{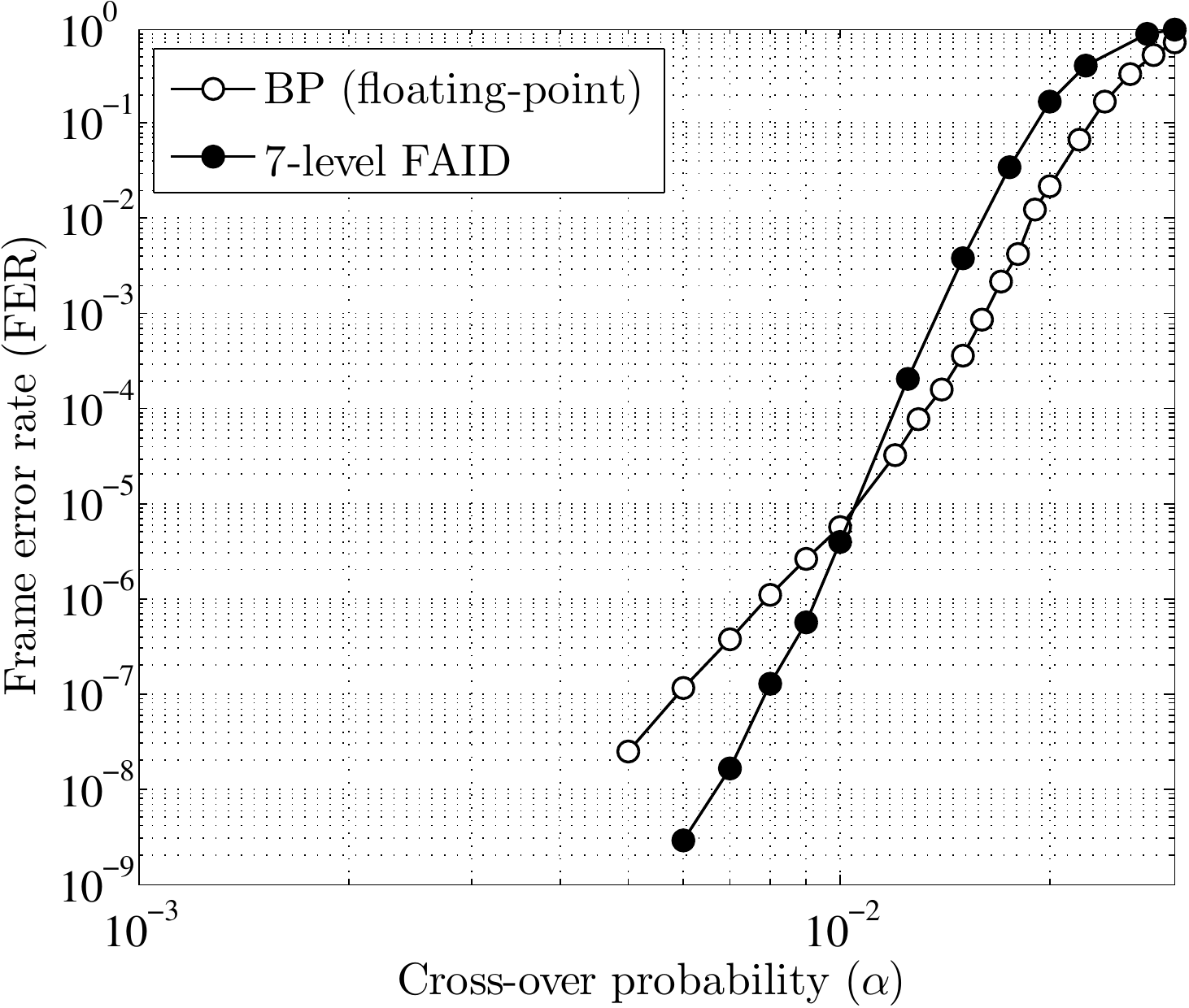}
\caption{Performance comparisons between the BP (floating-point), the the 7-level FAID defined by Table \ref{LUT244325} on the $(2388,1793)$ structured code.}
\label{PlotsQuasi2388}
\end{center}
\end{figure}

\begin{figure}[bhtp]
\begin{center}
\includegraphics[angle=0, width=3.5in]{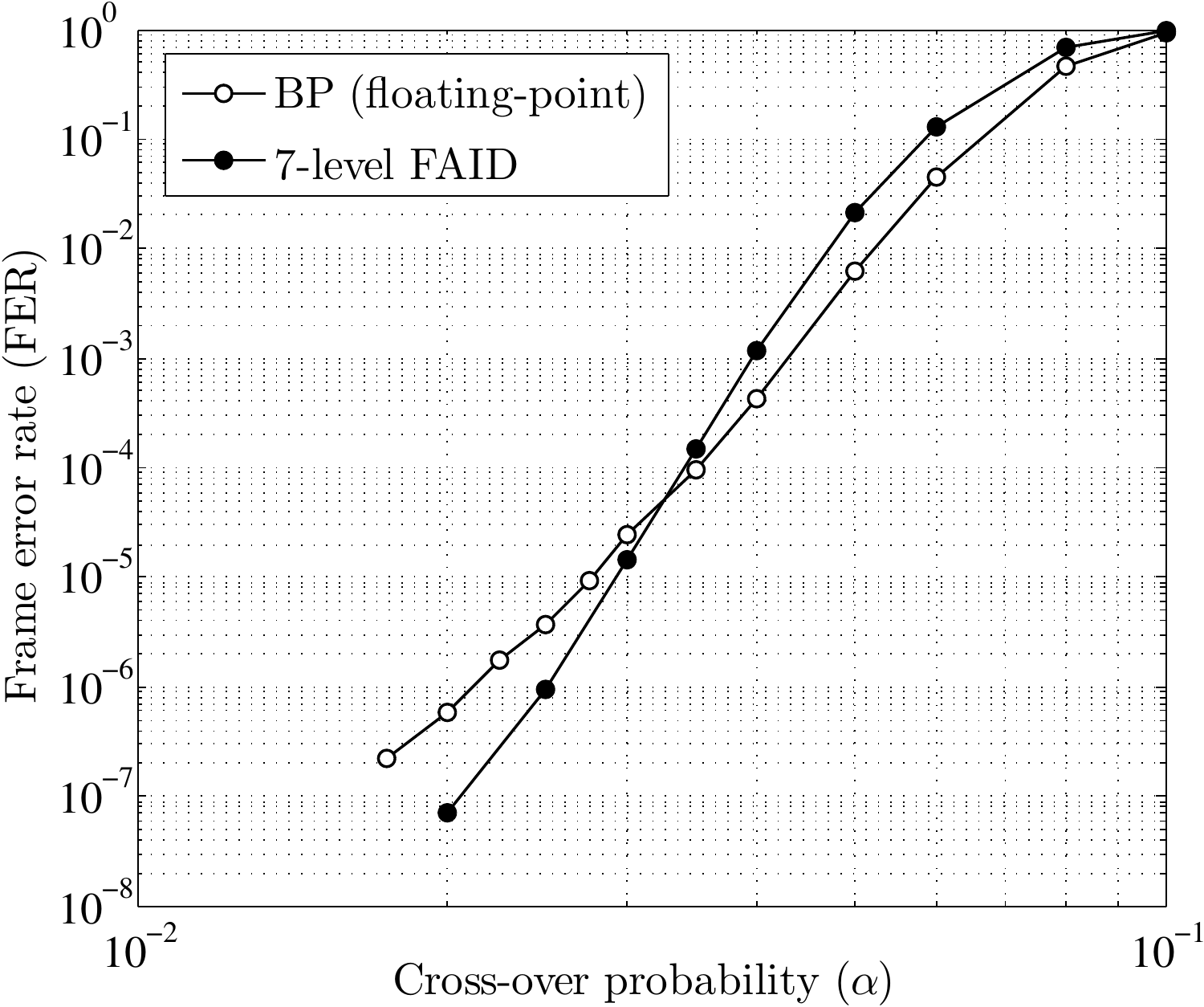}
\caption{Performance comparisons between the BP (floating-point) and the 7-level FAID defined by Table \ref{LUT1072079} on the $(502,252)$ quasi-cyclic code.}
\label{PlotsPEG502}
\end{center}
\end{figure}

\begin{figure}[bhtp]
\begin{center}
\includegraphics[angle=0, width=3.5in]{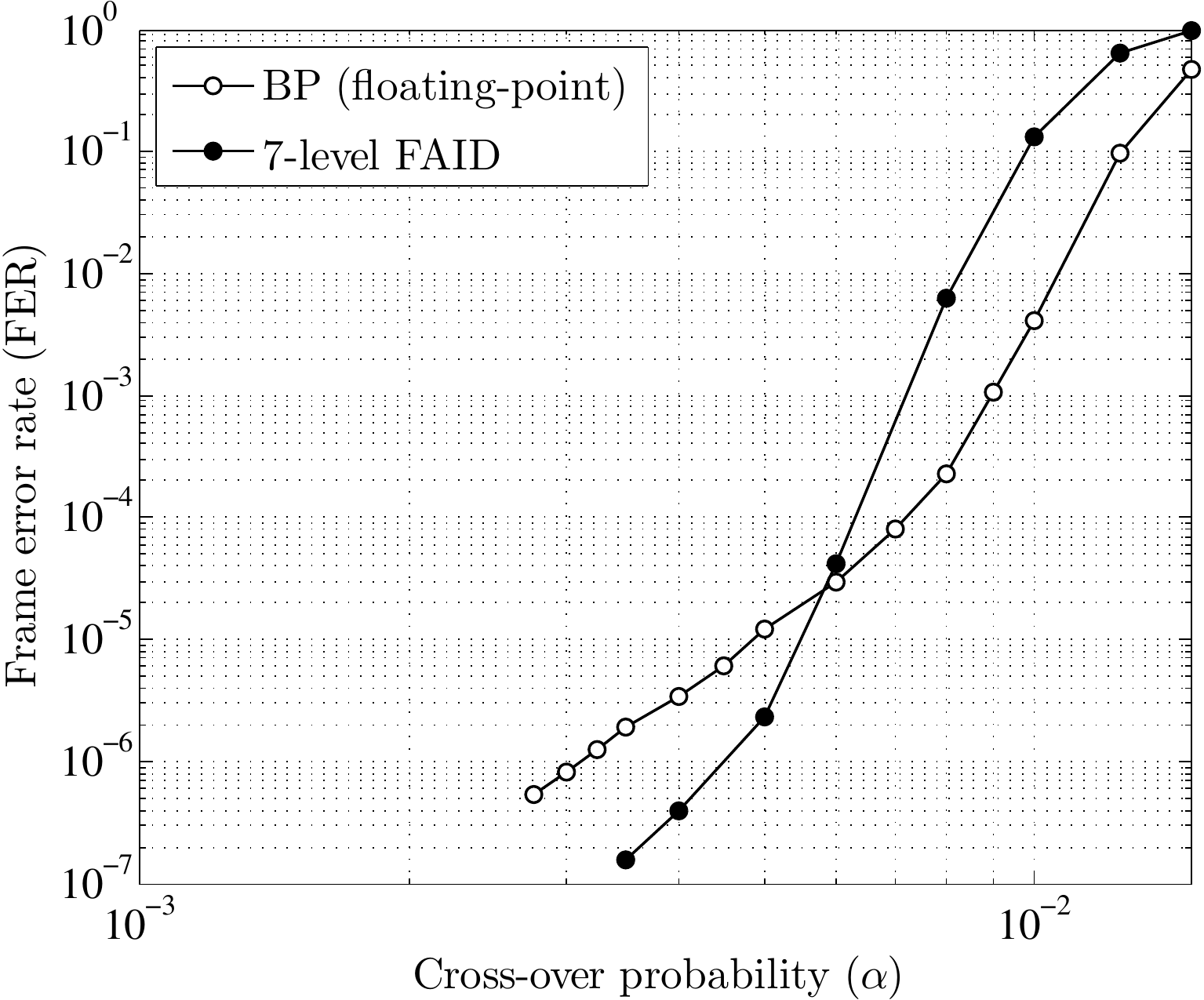}
\caption{Performance comparisons between the BP (floating-point) and the 7-level FAID defined by Table \ref{LUT244325} on the $(5184,4322)$ quasi-cyclic code.}
\label{PlotsQuasi5184}
\end{center}
\end{figure}

Figures \ref{PlotsQuasi2388}, \ref{PlotsQuasi5184}, and \ref{PlotsPEG502} show the frame error-rate (FER) performance comparisons versus the cross-over probability $\alpha$ between the particularly good 7-level (3-bit precision) FAIDs we identified and the BP (floating-point). Table \ref{LUT244325} defines the FAID used on the $(2388,1793)$ and the $(5184,4322)$ codes, while Table \ref{LUT1072079} defines the FAID used on the $(504,252)$ PEG-based code. Note that both are 7-level NLT FAIDs and all decoders were run for a maximum of 100 iterations. 

In all three codes, the 7-level FAIDs begin to surpass the BP at an FER$\simeq 10^{-5}$. Also notice the difference in the better slopes of the error floor for the 7-level FAIDs which can be attributed to their enhanced guaranteed error correction capability. For instance, all FAIDs used on the $(155,64)$ Tanner code in Fig. \ref{TannerPlots} guarantee a correction of 5 errors, whereas BP fails to correct several 5-error patterns. It must also be noted that the good 7-level FAIDs identified using our approach outperformed BP on several other tested codes as well. Therefore, the methodology is applicable to any column-weight-three code and provides to an extent ``universally'' good FAIDs, as they are all capable of surpassing BP on not just few but several codes. 

\section{Conclusions}
\label{sec:conclusions}

We introduced a new paradigm for finite precision iterative decoding of LDPC codes on the BSC. Referred to as FAIDs, the newly proposed decoders use node update maps that are much simpler than BP yet capable of surpassing the floating-point BP with only three bits of precision.  We described the general framework of FAIDs with focus on column-weight-three codes and provided examples of good 3-bit precision FAIDs. We also provided a general methodology to identify a set of ``universally" good FAIDs, one or several of which are potentially good for any given column-weight-three code. Our methodology is thus not code-specific but rather utilizes the knowledge of harmful topologies that could be present in a given code. The supporting numerical results show that it is possible to achieve a much superior error-rate performance in the error floor at a much lower complexity and memory usage than BP by using FAIDs. 

\appendix
\IEEEproof[Proof of Lemma 1] Assume that there exists an LT representation for such a $\Phi_v$, which is defined by assigning real values to the elements of the alphabet $\mathcal{M}$, the threshold set $\mathscr{T}$, and the set $\mathcal{Y}$. Since $\Phi_v(\mrm{-C},-L_1,L_2)=-L_1$, the inequality $L_1-L_2+\mrm{C}\geq T_1$ must hold. Also since $\Phi_v(-\mrm{C},0,L_1)=0$, we have $|\mrm{C} - L_1|<T_1$. Combining the two inequalities, we get $L_1-L_2+\mrm{C}>|\mrm{C}-L_1|$. 
Now, if $\mrm{C}>L_1$, then $2L_1>L_2$, and if $L_1\geq\mrm{C}$, then $L_1-L_2+\mrm{C}>L_1-\mrm{C}\Rightarrow 2\mrm{C}>L_2 \Rightarrow 2L_1>L_2$. But since $\Phi_v(-\mrm{C},0,-L_2)=-L_3$ and $\Phi_v(-\mrm{C},-L_1,-L_1)=-L_2$, we have $L_2+\mrm{C}>2L_1+\mrm{C}\Rightarrow L_2>2L_1$, which is a contradiction. 
\endIEEEproof

\section*{Acknowledgment}
The authors would like to thank Shashi Kiran Chilappagari for his contributions to this work.

\newpage

\begin{table}[bhtp]
\renewcommand{\arraystretch}{1.2}
	\caption{LUT for $\Phi_v$ of a 7-level FAID with $y_i=-\mrm{C}$}
	\label{LUT244325}
	\centering
\resizebox{9cm}{!}{
	\begin{tabular}{|c||c|c|c|c|c|c|c|}
	\hline
		\boldmath $m_{1}/m_{2}$   	& \boldmath$-L_3$	&\boldmath$-L_2$	& \boldmath$-L_1$ 	& \boldmath $0$ 	& \boldmath$+L_1$ 	& \boldmath $+L_2$ & \boldmath $+L_3$\\ \hline\hline
		\boldmath$-L_3$ 		& $-L_3$	&  $-L_3$	&  $-L_3$ 	&  $-L_3$	&$-L_3$	&  $-L_3$ &  $-L_1$\\ \hline
		\boldmath$-L_2$ 		& $-L_3$	&  $-L_3$	&  $-L_3$ 	&  $-L_3$	&$-L_2$	&  $-L_1$ &  $+L_1$\\ \hline
		\boldmath$-L_1$			& $-L_3$	&  $-L_3$ 	&  $-L_2$ 	&  $-L_2$	&$-L_1$	&  $-L_1$	  &  $+L_1$\\ \hline
		\boldmath $0$			& $-L_3$	&  $-L_3$	&  $-L_2$ 	&  $-L_1$	&$0$	&  $0$ &  $+L_1$\\ \hline
		\boldmath $+L_1$		& $-L_3$ 	&  $-L_2$	&  $-L_1$	&  $0$ 		&$0$	&  $+L_1$ &  $+L_2$\\ \hline
		\boldmath $+L_2$		& $-L_3$	&  $-L_1$ 	&  $-L_1$	&  $0$ 		&$+L_1$	&  $+L_1$ &  $+L_3$\\ \hline
		\boldmath $+L_3$		& $-L_1$	&  $+L_1$ 	&  $+L_1$ 	&  $+L_1$ 	&$+L_2$	&  $+L_3$ &  $+L_3$\\ \hline
	\end{tabular}
}
\end{table}

\begin{table}[bhtp]
\renewcommand{\arraystretch}{1.2}
	\caption{LUT for $\Phi_v$ of a 7-level FAID with $y_i=-\mrm{C}$}
	\label{LUT1072079}
	\centering
\resizebox{9cm}{!}{
	\begin{tabular}{|c||c|c|c|c|c|c|c|}
	\hline
		\boldmath $m_{1}/m_{2}$   	& \boldmath$-L_3$	&\boldmath$-L_2$	& \boldmath$-L_1$ 	& \boldmath $0$ 	& \boldmath$+L_1$ 	& \boldmath $+L_2$ & \boldmath $+L_3$\\ \hline\hline
	\boldmath	$-L_3$	&	$-L_3$	&	$-L_3$	&	$-L_3$	&	$-L_3$	&	$-L_3$	&	$-L_3$	&	0	\\  \hline
\boldmath	$-L_2$	&	$-L_3$	&	$-L_3$	&	$-L_3$	&	$-L_2$	&	$-L_2$	&	$-L_1$	&	$L_1$	\\  \hline
\boldmath	$-L_1$	&	$-L_3$	&	$-L_3$	&	$-L_2$	&	$-L_2$	&	$-L_1$	&	0	&	$L_2$	\\  \hline
\boldmath	0	&	$-L_3$	&	$-L_2$	&	$-L_2$	&	$-L_1$	&	0	&	$L_1$	&	$L_2$	\\  \hline
\boldmath	$L_1$	&	$-L_3$	&	$-L_2$	&	$-L_1$	&	0	&	0	&	$L_1$	&	$L_2$	\\  \hline
\boldmath	$L_2$	&	$-L_3$	&	$-L_1$	&	0	&	$L_1$	&	$L_1$	&	$L_2$	&	$L_3$	\\  \hline
\boldmath	$L_3$	&	0	&	$L_1$	&	$L_2$	&	$L_2$	&	$L_2$	&	$L_3$	&	$L_3$	\\  \hline
	\end{tabular}
}
\end{table}  

\begin{thebibliography}{99}

\bibitem{gallager} R. G. Gallager, {\it Low Density Parity Check Codes}. Cambridge, MA: M.I.T. Press, 1963. 
\bibitem{pearl} J. Pearl, {\it Probablisitic Reasoning in Intelligent Systems}. San Francisco, CA: Kaufmann, 1988.




\bibitem{richardsonerrorfloor} T. Richardson, ``Error floors of LDPC codes,'' {\it Proc. Allerton Conf. on Commun., Control and Computing}, 2003.
%
%

\bibitem{yedidia} J. S. Yedidia, W. T. Freeman, and Y.Weiss, ``Constructing free energy approximations and generalized belief propagation algorithms,'' {\it IEEE Trans. Inf. Theory}, vol. 51, pp. 2282-–2312, Jul. 2005. 




\bibitem{milos} M. Ivkovic, S. K. Chilappagari, and B. Vasic, ``Eliminating trapping sets in low-density parity-check codes by using Tanner graph covers,'' {\it IEEE Trans. Inf. Theory}, vol. 54, no. 8, pp. 3763--3768, Aug. 2008. 

\bibitem{DungLatinSquare} D. V. Nguyen, S. K. Chilappagari, M. W. Marcellin, and B. Vasic, ``On the construction of structured LDPC Codes free of small trapping sets,'' {\it IEEE Trans. Inf. Theory}, vol. 58, no. 4, pp. 2280--2302, Apr. 2012.  


\bibitem{richardson} T. Richardson and R. Urbanke, ``Capacity of low-density parity-check codes under message-passing decoding,'' {\it IEEE Trans. Inf. Theory}, vol 47, pp. 599--618, Feb. 2001. 

\bibitem{chen} J. Chen, A. Dholakia, E. Eleftheriou, M. Fossorier, and X.-Y. Hu, ``Reduced-complexity decoding of LDPC codes,'' {\it IEEE Trans. Commun.}, vol. 53, no. 8, pp. 1288--1299, Aug. 2005.

\bibitem{Fossorier} M. Fossorier, M. Mihaljevic, H. Imai, ``Reduced complexity iterative decoding of low-density parity check codes based on belief propagation,'' {\it IEEE Trans. Commun.}, vol. 47, no. 5, pp. 673--680, May 1999.

\bibitem{leethorpe} J. Lee, J. Thorpe, ``Memory-efficient decoding of LDPC codes,'' {\it Proc. Int. Symp. Inform. Theory}, pp. 459--463, Sept. 2005.

\bibitem{kurkoski} B. Kurkoski and H. Yagi, ``Quantization of binary-input Discrete Memoryless Channels with applications to LDPC decoding,'' {\it submitted to IEEE Trans. Inf. Theory} [Online]. Available: http://arxiv.org/abs/1107.5637 


\bibitem{Planjery_ElecLetters_2011} S. K. Planjery, D. Declercq, L. Danjean, and B. Vasic, ``Finite alphabet iterative decoders for LDPC codes surpassing floating-point iterative decoders,'' {\it IET Electron. Lett.}, vol. 47, no. 16, Aug. 2011. 

\bibitem{PlanjeryITA2010}S.K. Planjery, S.K. Chilappagari, B. Vasic, D. Declercq, L. Danjean, ``Iterative decoding beyond belief propagation,''{\it Proc. Inform. Theory and App. Workshop}, Jan. 2010.

\bibitem{LudoITW2011} L. Danjean, D. Declercq, S. K. Planjery, and B. Vasic, ``On the selection of finite alphabet iterative decoders for LDPC codes on the BSC,'' {\it Proc. IEEE Inform. Theory Workshop}, pp. 345--349, Oct. 2011. 


\bibitem{DolecekDec} L. Dolecek, P. Lee, Z. Zhang, V. Anantharam, B. Nikolic, and M. Wainwright,``Predicting Error Floors of Structured LDPC Codes: Deterministic Bounds and Estimates,''{\it IEEE J. Sel. Areas Commun.}, vol. 27, pp. 908--917, Aug. 2009. 

\bibitem{varnica}N. Varnica, M. Fossorier, and A. Kavcic, ``Augmented belief propagation decoding of low-density parity check codes,'' {\it IEEE Trans. Commun.}, vol. 55, no. 7, pp. 1308--1317, Jul. 2007.


\bibitem{Wesel} A. I. Vila Casado, M. Griot, and R. D. Wesel,``LDPC decoders with informed dynamic scheduling,'' {\it IEEE Trans. Commun.}, vol. 58, no. 12, pp. 3470--3479, Dec. 2010. 

\bibitem{ywang} Y. Wang, J. S. Yedidia, and S. C. Draper, ``Multi-stage decoding of LDPC codes,'' {\it Proc. IEEE Int. Symp. Inform. Theory}, pp. 2151--2155, Jul. 2009.

\bibitem{laendner} S. Laendner, and O. Milenkovic, ``Algorithmic and combinatorial analysis of trapping sets in structured LDPC codes ,'' {\it Proc. Int. Conf. Wireless Networks, commun., and mobile commun.}, pp. 630--635, Jun. 2005.

\bibitem{YHan} Y. Han, and W. Ryan, ``Low-floor decoders for LDPC codes,'' {\it IEEE Trans. Commun.}, vol. 57, no. 6, pp. 1663--1673, 2009.

\bibitem{zhangerrorfloor} Z. Zhang, L. Dolecek, B. Nikoli\'c, V. Anantharam, and M. Wainwright, ``Lowering LDPC error floors by postprocessing,'' {\it Proc. IEEE Global Telecommun. Conf.}, pp. 1--6,  Dec. 2008.

\bibitem{bannihasemi} F. Zarkeshvari, A. H. Banihashemi, ``On implementation of min-sum and its modifications for decoding LDPC codes,'' {\it IEEE Trans. Commun. lett.}, vol. 53, no. 4, pp. 549--554, Apr. 2005.

\bibitem{DolecekQuantization} Z. Zhang, L. Dolecek, B. Nikoli\'c, V. Anantharam, and M. Wainwright, ``Design of LDPC decoders for improved low error rate performance: quantization and algorithm choices,'' {\it IEEE Trans. Commun.}, vol. 57, no. 11, pp. 3258--3268, Nov. 2009.

\bibitem{shashiICC} S. K. Chilappagari, S. Sankaranarayanan, and B. Vasic, ``Error floors of LDPC codes on the binary symmetric channel,'' {\it Proc. IEEE Int. Conf. on Commun.}, pp. 1089--1094, Jun. 2006.

\bibitem{Kuperberg_symmetries_of_plane_partitions} G. Kuperberg, ``Symmetries of plane partitions and the permanent-determinant method,'' {\it J. Comb. Theory A}, 68, pp. 115--151, 1994.

\bibitem{shashiITpaper} S. K. Chilappagari, and B. Vasi\'c, ``Error-correction capability of column-weight-three LDPC codes,'' {\it IEEE Trans. Inf. Theory}, vol. 55, no. 5, pp. 2055--2061, May 2009.

\bibitem{ontology} B. Vasic, S. K. Chilappagari, D. V. Nguyen, and  S. K. Planjery, ``Trapping set ontology,'' {\it Proc. Allerton Conf. on Commun., Control, and Computing}, pp. 1--7, Sept. 2009.

\bibitem{planjeryISIT2010} S. K. Planjery, D. Declercq, S. K. Chilappagari, and B. Vasi\'{c}, ``Multilevel decoders surpassing belief propagation on the binary symmetric channel," {\it Proc. IEEE Int. Symp. Inform. Theory}, pp. 769--773, Jun. 2010.

\bibitem{Declercq_ISTC_2010} D. Declercq, L. Danjean, E. Li, S. K. Planjery, and B. Vasic, ``Finite alphabet iterative decoding (FAID) of the (155,64,20) Tanner code,'' {\it Proc. Int. Symp. Turbo Codes Iter. Inform. Process.}, pp. 11--15, Sep. 2010.

\bibitem{PEGpaper} X. Y. Hu, E. Eleftheriou, and D. M. Arnold, ``Regular and irregular progressive edge-growth tanner graphs,'' {\it IEEE Trans. Inf. Theory}, vol. 51, no. 1, pp. 386--398, Jan. 2005. 
\end{thebibliography}
\end{document}